\documentclass{aastex63}

\usepackage{amsmath}

\shorttitle{Helical jet of the blazar S5 0716+714}
\shortauthors{Butuzova M.S.}

\published{Astronomy Reports, Volume 62, Issue 2, pp.116-122, 2018,
 DOI:  10.1134/S1063772918020038}

\begin{document}

\title{Geometrical and Kinematic Parameters of the Jet of the Blazar
S5 0716+71 \\in a Helical-Jet Model}

\correspondingauthor{Marina Butuzova}
\email{mbutuzova@craocrimea.ru}

\author[0000-0001-7307-2193]{Marina S. Butuzova}
\affiliation{Crimean Astrophysical Observatory of RAS, 298409, Nauchny, Russia}

\begin{abstract}

Periodic variations of the position angle of the inner jet of the blazar S5 0716+71 suggest a helical structure for the jet. The geometrical parameters of a model helical jet are determined. It is shown that, when the trajectories of the jet components are non-ballistic, the angle between their velocity vectors and the line of sight lies in a broader interval than is the case for ballistic motions of the components, in agreement with available estimates.
The contradictory results for the apparent speeds of components in the inner and outer jet at epochs 2004 and 2008$-$2010 can be explained in such a model. The ratio of the apparent speeds in the inner and outer jet is used to derive a lower limit for the true speed of
the components $\left(\beta>0.999 \right)$ and to determine the pitch angle of the helical jet $\left(p=5.5^\circ \right)$. 
The derived
parameters can give rise to the conditions required to observe high speeds (right to 37$c$) for individual jet
components.

\end{abstract}

\keywords{blazar, S5 0716+714, helical jet}

\section{Introduction} 
\label{sec:intro}

The BL~Lac object S5~0716+71 is a highly variable source on time scales of both about a day \citep{Montagni06, Poon09,Volvach09, Gorshkov11, Gorshkov11b} and several years \citep{Dai13,Volvach12, Bychkova15}.
Its redshift is not precisely known; estimates of$z$ based on observations of nearby galaxies range from $z=0.26$ to $z\geq0.52$ \citep{Wagner96, Bychkova06, Sbarufatti05}.
It has been intensely observed over the entire
electromagnetic spectrum \citep[see, e.g.][]{Wagner96, Liao14}.
Very Long Baseline Interferometry (VLBI) observations of
S5~0716+71 have been conducted over more than 10~years \citep{Bach05,Britzen09, Rani15}, for example as part of the MOJAVE program \citep{Lister13}.
VLBI observations at 5, 8.4, 15, and
22~GHz during 1992.74---2001.2 revealed variations
in the position angle of the inner jet at distances up to
1~milliarcsecond (mas) from the core with a period of $7.4\pm1.5$~yrs and an amplitude of $3.5^\circ$ \citep{Bach05}.
Analysis of 15~GHz VLBI data for 1994.5---2011.5 showed the
period of variations of the position angle of the inner
part of the jet (closest to the core), $\text{PA}_\text{in}$, 
to be 10.9~yrs, with an amplitude of $11^\circ$ \citep{Lister13}.
Periodic variations of $\text{PA}_\text{in}$ can be most simply explained if the jet has a helical shape \citep{Bach05,Lister13}.

This hypothesis has been widely applied in interpretations
of the observed properties of the jets of
active galactic nuclei (AGNs). For example, a helical
jet shape arising due to precession of the central black
hole has been used to explain the apparent motions
of superluminal components and long-period brightness
variations in the AGNs OJ~287 \citep{Abraham00}, BL~Lac \citep{CapAbr13}, 3C~120 \citep{CapAbr04}, and 3C 279 \citep{AbrCar98}.
The shape of the spectral energy distribution of the blazar Mrk~501 and its time variations have also been interpreted in
terms of a helical jet model \citep{VilRait99}.
A helical jet could form as a result of the development of magnetohydrodynamical instability arising in the jet flow \citep[see,e.g.,][]{HardeeNorman88}.
Spiral structure could also form due to the
twisting of a magnetic rope in and around a conical
jet \citep{MeierKU01}, and the plasma near a magnetized accretion disk follows helical trajectories \citep{CamKrock92}.
Returning to the blazar S5~0716+71, it has been proposed that a shock moving along a curved, possibly helical, trajectory
can provide a good explanation for the behavior of its
gamma-ray, optical, and radio flares \citep{Rani15,Larionov13, Rani13}.

In the study presented here, we have used the
period and amplitude of variations of the inner-jet
position angle PA$_\text{in}$ \citep{Lister13} to derive the geometrical
parameters of the helical curve formed by the jet
components (Section~2). In Section~3, we use these
parameters to derive intervals of possible values for
the angle between the line of sight and the velocity
vectors of the jet components for the cases of ballistic
and non-ballistic motion. We show in Section 4
that it is possible to reconcile the different detected
apparent component speeds for the inner and outer jet
of S5 0716+71 presented by citet{Rastorgueva11} for epoch 2004 and by \citet{Rani15} for epoch
2008$-$2010 only if the component motions are non-ballistic.
In this case, the condition for high apparent
jet-component speeds reaching $37c$, as is observed for
S5~0716+71 \cite{Rani15}, is realized. It is also possible to
determine a lower limit on the true speed of the
jet components. The main results of this study are
summarized in Section~5.

It is important that our conclusions do not depend
on the processes leading to the helical shape of the jet
or the physical nature of the jet components. The reasoning
and conclusions presented here can be applied
to various objects that we can treat as components
of a jet. For example, if radiation is produced by
the entire jet, a jet component could correspond to a
volume element in the jet. If distinct parts of a stream
radiate (plasmoids, shock fronts where electrons are
accelerated and then radiate), we can also consider
these regions as jet components, without specifying
their nature.

\section{Determination of the geometrical parameters of a helical jet}

\begin{figure}
    \centering
    \includegraphics[scale=0.55]{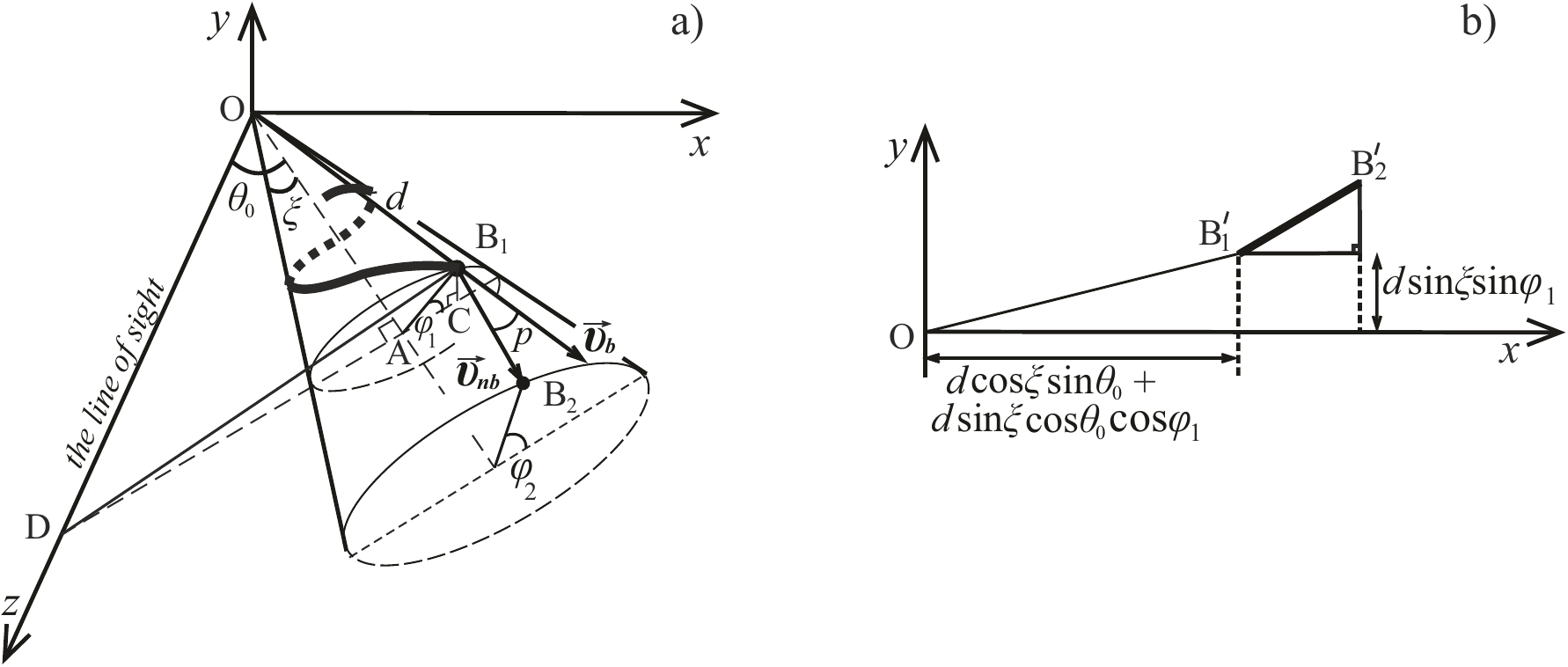}
    \caption{(a) Schematic of the geometrical parameters of a
helical jet. The jet is shown by the bold curve (a section
of the jet located on the opposite side of the cone is shown
dotted). At the time shown in this figure, the component
B$_1$ is located a distance $d$ from the vertex of the cone.
The velocity vectors for ballistic ($\boldsymbol{\upsilon}_b$) and non-ballistic ($\boldsymbol{\upsilon}_{nb}$) motions are shown. (b) Projection of the schematic (a) onto the celestial sphere. The segment $\text{B}^\prime_1 \text{B}^\prime_2$ is the
projection of the trajectory of a component moving with
speed $\boldsymbol{\upsilon}_{nb}$ over a time $dt$.}
    \label{fig:fig1}
\end{figure}

Let us suppose that a jet component is located
on the surface of a cone and follows a helical curve
beginning at some distance from the vertex of the
cone (Fig.~\ref{fig:fig1}a). The vertex of this cone may not
coincide with the position of the central supermassive
black hole or the position of the physical beginning of
the jet. The latter probably has a parabolic shape, but
the size of this region is substantially smaller than the
distances considered in this study.

We introduce a coordinate system with its origin
at the cone vertex. The OZ axis is directed along
the line of sight and the OX axis along the projection
of the cone axis OA onto the plane of the sky. The
opening angle of the cone and the angle between the
cone axis and the line of sight are denoted $\xi$ and $\theta_0$, respectively. All information about angles is given in
degrees. \citet{Lister13} determined PA$_\text{in}$ as the mean
flux-density-weighted position angles of all the jet
features located at distances of $0.15-1$~mas from the
core. In contrast to this, we will simply consider the
position angles of components measured at the time
when they reach a distance $d$ from the cone vertex.
Note that the concept of the ``inner jet'' refers to that
part of the jet located out to some distance from the
VLBI core, usually specified in various studies to be
$1-2$~mas. Another quantity specifying the position
of a component on the surface of the cone is the
azimuth angle $\varphi$, defined as the angle between the
OXZ and OAB$_1$ planes. As the jet material moves
outward, the angle $\varphi$ for each successive component
reaching the distance $d$ differs from that for the previous
component, leading to variations in PA$_\text{in}$.
When $\theta_0<\xi$, the line of sight is inside the helical path
formed by the jet. In this case, the inner-jet position
angle PA$_\text{in}$ varies by $360^\circ$ over the period of variations of $\varphi$, which does not agree with the observations of \citet{Bach05, Lister13}. Therefore, we have restricted
our consideration to the case $\theta_0<\xi$.

\begin{figure}
    \centering
    \includegraphics[scale=0.7]{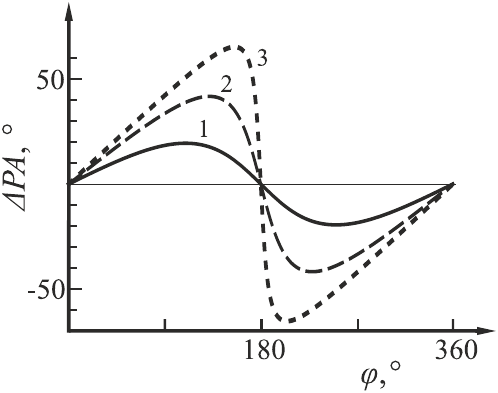}
    \caption{Theoretical deviation of the position angle of the
inner jet from its mean value over one variation period for
$\varphi$, for $\theta_0/\xi=3$ (1), 1.5 (2), and 1.1 (3).}
    \label{fig:fig2}
\end{figure}

The mean position angle of the inner jet (PA$_0$)
corresponds to the position of a component in the
OXZ plane. Deviations of the measured values for PA$_\text{in}$ from PA$_0$, i.e., the angle between the OX axis
and the radius vector of a component projected onto
the plane of the sky (OB$_1$), is denoted $\Delta\text{PA}$.
As is shown in Fig.~\ref{fig:fig1}b, an expression for $\Delta\text{PA}$ in terms of $\varphi$, $\theta_0$, and $\xi$ can straightforwardly be found from the projections of OA, OB$_1$, and B$_1$C onto the plane of
the sky:
\begin{equation}
%\begin{split}
    \tan \Delta PA=\frac{\sin \xi \sin \varphi}{\cos \xi \sin \theta_0+\sin \xi \cos \theta_0 \cos \varphi} \approx 
    \frac{\sin \varphi}{\theta_0/\xi+\cos \varphi},
    \label{eq:tanDPA}
%\end{split}
\end{equation}
where we have approximated that the angles $\theta_0$ and $\xi$
are small. The observed position angle is then $\text{PA}_\text{in}=\text{PA}_0+\Delta\text{PA}$. As can be seen from the right-hand side of (\ref{eq:tanDPA}), when $\theta_0$ and $\xi$ are small, PA$_\text{in}$ depends on the ratio $\theta_0/\xi$. For definiteness, we will further assume
that the helical jet is winding in the direction that
leads to an increase in $\varphi$ as each successive component
appears at the distance $d$. Figure~\ref{fig:fig2} presents the
variations of PA$_\text{in}$ over one period calculated using the
exact formula~(\ref{eq:tanDPA}), for various values of the ratio $\theta_0/\xi$. As this ratio increases, the amplitude of the variations
of PA$_\text{in}$ and the degree of asymmetry of the variation
curve both decrease.

It follows from (\ref{eq:tanDPA}) that
\begin{equation}
    \varphi=\pm\arccos\left(-\tan \xi \cot \theta_0 \right),
    \label{eq:fi}
\end{equation}
and PA$_\text{in}$ has maximum [$+$ in (\ref{eq:fi})] and minimum
[$-$ in (\ref{eq:fi})] values. Substituting (\ref{eq:fi}) into (\ref{eq:tanDPA}) and making the substitution $\sin \varphi=\sqrt{1-\cos^2\varphi}$ yields the amplitude
of the variations of PA$_\text{in}$:
\begin{equation}
    \text{PA}_\text{max}=\arctan\left[\sin \xi \left(\cos^2 \xi \sin^2\theta_0-\sin^2\xi\cos^2\theta_0 \right)^{-1/2} \right]
    \label{eq:PAmax}\approx\left[\left(\frac{\theta_0}{\xi} \right)^2-1 \right]^{-1/2},
\end{equation}
which, like $\Delta\text{PA}$, depends only on the ratio of $\theta_0$ and $\xi$ when these angles are small. To determine $\theta_0/\xi$ using (\ref{eq:tanDPA}), we used the value PA$-\text{max}=11^\circ$ \citep{Lister13}, based
on a more extensive series of observations with better
angular resolution than in \citep{Bach05}. Then,
\begin{equation}
    \theta_0/\xi\approx5.3.
    \label{eq:ratio}
\end{equation}
Various estimates of the angle of the jet of
S5~0716+71 to the line of sight based on the observed
apparent speeds of VLBI components have yielded $\theta_0$ values from $0.5^\circ$ to $12^\circ$ \citep{Bach05,Britzen09,Rastorgueva11}. However, these estimates were carried out using observations
covering limited time intervals, and depend on the jet
model considered. Therefore, we focused on the value
obtained in \citep{Pushkarev09}, where $\theta_0$ values were determined
from the maximum apparent speeds of components
displaying radial motion without acceleration, observed
in the MOJAVE project over no fewer than five
epochs. The value obtained for S5~0716+71 was $\theta_0=5.3^\circ$ \citep{Pushkarev09}. A very similar value for $\theta_0$ was obtained
by \citet{Savolainen10}. It follows from (\ref{eq:ratio}) that $\xi\approx1^\circ$. Hence, the
opening angle of the jet, that is, the angle inside of
which jet components can be observed, is $2\xi=2^\circ$. The opening angle deduced from analysis of VLBI
maps of the blazar S5~0716+71 is $1.6^\circ$ \citep{Pushkarev09}. Thus, we can see a good consistency between these estimates
of the jet opening angle derived using independent methods.

\section{Angle between the velocity vector of a jet component and the line of sight}

\begin{figure}
    \centering
    \includegraphics[scale=0.5]{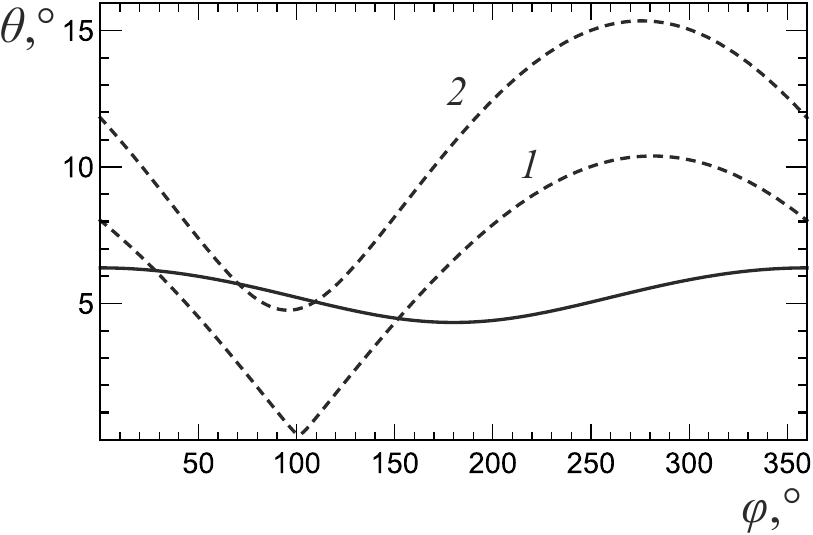}
    \caption{Dependence of the direction of the motion of
jet components relative to the line of sight ($\theta$) on the
azimuthal angle $\varphi$ for ballistic motion (solid curve) and
non-ballistic motion (dashed curves) with $p = 5^\circ$ (1) and
$p = 10^\circ$ (2).}
    \label{fig:fig3}
\end{figure}

In the case of ballistic motion, the velocity vector of
a component ($\boldsymbol{\upsilon}_b$) located a distance $d$ from the cone
vertex is directed along the cone generatrix (Fig.~\ref{fig:fig1}).
The angle $\theta_b$ between $\boldsymbol\upsilon_b$ and the line of sight can be found from the triangle OB$_1$D.
The side OD can
be found from the triangle OAD together with the
relations $\angle\text{DOA}=\theta_0$ and $\text{OA}= d \cos\xi$, and the side
B$_1$D can be found from the triangle AB$_1$D with the
relations $\angle\text{DAB}_1=180^\circ-\varphi$, AB$_1 = d \sin \xi$, AD$ =d \cos \xi \tan \theta_0$. The final expression for $\theta_b$ does not depend on $d$:
\begin{equation}
    \cos \theta_b=\cos \xi \cos \theta_0-\sin \xi \sin \theta_0 \cos \varphi.
    \label{eq:costhb}
\end{equation}
The value of $\theta_b$ decreases non-uniformly from $\theta_0+\xi=6.3^\circ$ to $\theta_0-\xi=4.3^\circ$ as $\varphi$ varies from 0$\circ$ to 180$\circ$, and grows symmetrically back to 6.3$^\circ$ as $\varphi$ varies from 180$^\circ$ to 360$\circ$ (solid curve in Fig.~\ref{fig:fig3}).

In the case of non-ballistic motion, the velocity vector of a component is directed at a pitch angle $p$ to the cone generatrix.
The angle between the line of sight and this velocity vector $\upsilon_{nb}$ can be obtained from the expression
\begin{equation}
    \sin \theta_{nb}=\text{B}^\prime_1\text{B}^\prime_2/\text{B}_1\text{B}_2,
    \label{eq:sinthnb}
\end{equation}
where $\text{B}^\prime_1\text{B}^\prime_2$ is the projection onto the plane of the sky of the segment $\text{B}_1\text{B}_2=|\upsilon_{nb}|\cdot \Delta t$, traversed by the component over a time $\Delta t$.
Figure~\ref{fig:fig1}b presents the
segment $\text{B}^\prime_1\text{B}^\prime_2$ together with the coordinates of the point $\text{B}^\prime_1$.
The coordinates of the point $\text{B}^\prime_2$ can be found
straightforwardly by replacing $\varphi_1$ with $\varphi_2$ and $d$ with $\left(d+\Delta d \right)$, where $\Delta d=\upsilon _{nb}\Delta t \cos p$ is the increase in the distance from the cone vertex to the component considered over a time $\Delta t$. As can be seen in Fig.~\ref{fig:fig1}b, $\text{B}^\prime_1\text{B}^\prime_2$ can easily be found as
\begin{equation}
    \text{B}^\prime_1\text{B}^\prime_2=\sqrt{f_a^2+f_b^2},
    \label{eq:b1b2}
\end{equation}
where
\begin{equation}
f_a=(d+\Delta d)\sin \xi \sin \varphi_2-d \sin \xi \sin \varphi_1,
    \label{eq:fa}
\end{equation}
\begin{equation}
    f_b=(d+\Delta d)(\sin \xi \cos \theta_0 \cos \varphi_2+\cos \xi \sin \theta_0)-d(\sin \xi \cos \theta_0 \cos \varphi_1+\cos \xi \sin \theta_0),
    \label{eq:fb}
\end{equation}
and $\varphi_1$ and $\varphi_2$ are the azimuth angles of the points
B$_1$ and B$_2$. Knowing the change in the azimuth angle
\begin{equation}
    \Delta \varphi=\frac{\upsilon_{nb}\Delta t \sin p}{(d+\Delta d)\sin \xi},
    \label{eq:dfi}
\end{equation}
taking this to be small over the time $\Delta t$, and using the
relation $\varphi_2=\varphi_1+\Delta \varphi$, the expression (6) acquires
the form
\begin{equation}
    \sin \theta_{nb}=\sqrt{g_a^2+g_b^2},
    \label{eq:thnb}
\end{equation}
where
\begin{equation}
    g_a=\cos p \sin \xi \sin \varphi_1+\sin p \cos \varphi_1,
    \label{eq:ga}
\end{equation}
\begin{equation}
    g_b=\cos p (\cos \xi \sin \theta_0+\sin \xi \cos \theta_0 \cos \varphi_1)-\sin p \cos \theta_0 \sin \varphi_1.
    \label{eq:gb}
\end{equation}
If $\varphi_2=\varphi_1-\Delta \varphi$, the signs of the terms $\sin p \cos \varphi_1$ and $\sin \xi \cos \theta_0 \cos \varphi_1$ change. However, our conclusions do not depend on the sign of $\varphi_2-\varphi_1$.

Analyzing $\theta_{nb}\left(\varphi \right)$ using formulas~(\ref{eq:sinthnb})---(\ref{eq:gb}) with various values of $p$, we found that this quantity lies in the interval $\theta_{nb}\approx p\pm 5^\circ$. Figure~\ref{fig:fig3} shows the variations of $\theta_{nb}$ for the cases $p = 5^\circ$ and 10$^\circ$ (dashed curves).

\section{Apparent and physical speeds of the jet components in S5~0716+71}

The results of different studies of the kinematics
of the jet of the blazar S5~0716+71 are contradictory.
Based on observations at 5, 8.4, 15, and 22~GHz
obtained in 1992.73---2001.17, \citet{Bach05} report
the presence of components moving rapidly in the
radial direction (apparent speeds $\beta_\text{app}=5c-16c$),
whereas more typical values for BL~Lac objects are
$\beta_\text{app}\leq5c$ \citep{GabuzdaPush2000}.
On the other hand, observations
from 1992.73---2006.32 did not exhibit any appreciable
outward component motions, instead indicating
the presence of stationary components \citep{Britzen09}.
That is, the jet components were located at roughly the same
distances from the core during this entire interval, and
any variations in their position angles were within the
uncertainties. An analysis of data obtained at 43 and
86~GHz during 2008---2010 indicated the presence of
features in the inner jet (within 2~mas from the core)
with apparent speeds up to 10$c$, which was lower
than the values of $\beta_\text{app}$ detected farther from the core
($>20c$) \citep{Rani15}.
However, a contradictory result was
obtained by \citet{Rastorgueva11} using data at 1.6,
5, 22, 43, and 86~GHz from 2004: components in
the inner jet (to 1~mas from the core) moved rapidly,
with speeds up to $\beta_\text{app}\approx 20 c$, while the maximum
apparent speed observed for components in the outer
jet was 10$c$.

Let us consider the results indicated above in our
model of a helical jet. According to the 2004 data \citep{Rastorgueva11},
the position angle of the inner jet was close to its
maximum value, $\approx 25^\circ$, while the position angle of
the outer jet was close to its minimum value, 15$^\circ$.
We found from (\ref{eq:fi}) that the model azimuth angles
of the inner and outer jet features are $\varphi_\text{fast}\simeq 101^\circ$ and
$\varphi_\text{low}\simeq259^\circ$, respectively.
\citet{Rani15} considered data obtained from September 2008 through October
2010. By this time, the inner components observed
in 2004 had moved to the outer part of the jet, and
the inner jet is formed by new components whose
position angles have their minimum value \citep[see Fig.~8 in][]{Lister13}.
Thus, the azimuthal angles of the inner
and outer jet components in 2008---2010 are $\varphi_\text{slow}\simeq259^\circ$ and $\varphi_\text{fast}\simeq101^\circ$, respectively.
The results of an
analysis of the kinematics of the jet components in
the interval 1992.72---2001.17 indicate that features in
the inner jet moved systematically more slowly than
those in the outer jet \citep{Bach05}.
Since, as in 2008---2010,
PA$_\text{in}$ had its minimum value in 1997---1998 \citep[see Fig.~8 in][]{Lister13}, this result is also consistent with the conclusion
that the rapidly moving components observed
in both 2004 and in 2008---2010 have azimuth angles
of 101$^\circ$, while slowly moving components have
azimuth angles of 259$^\circ$.
Note that the physical
speed of the components $\beta$ (in units of the speed of
light) must be roughly the same in order for the helical
structure of the jet to exist over a fairly long time.
With a constant $\beta$, the apparent speed
\begin{equation}
    \beta_\text{app}=\frac{\beta \sin \theta}{1-\beta \cos \theta}
    \label{eq:bapp}
\end{equation}
depends on the angle $\theta$ between the velocity vector of
the component and the line of sight. With the indicated
azimuth angles, it follows from (\ref{eq:costhb}) that, in the
case of ballistic motion of components with low and
high apparent speeds directed at the same angle to
the line of sight, $\theta_\text{fast}=\theta_\text{slow}\approx8^\circ$.
Differences in the
apparent speeds must then be due to differences in the
physical speeds of the components, which contradicts
our assumption that $\beta$ is constant, and we believe that
such variations in $\beta$ do not have a natural physical
explanation. In the case of non-ballistic motion with
a pitch angle of $p\approx5^\circ$, we find from (\ref{eq:sinthnb})---(\ref{eq:gb}) that the angles to the line of sight for fast and slow components
are $\theta_\text{fast} \approx 3^\circ$ and $\theta_\text{slow} \approx 13^\circ$, respectively. This can explain the difference in the apparent speeds
of components having the maximum and minimum
position angles. Consequently, the seemingly contradictory
results of the kinematic analyses of components
in the jet of the blazar S5~0716+71 obtained
by \citet{Rani15} and \citet{Rastorgueva11} can only be explained in a natural way
in the case of non-ballistic motions of components, with different azimuthal angles for the inner and outer features at different observing epochs. Note that a
number of studies have suggested the existence of
non-ballistic trajectories for component motions \citep[see, e.g.,][]{Britzen09, Rani15}. However, in this case, $\beta_\text{app}$ depends
not only on $\varphi$, but also on $p$. Figure~ref{fig:fig4} shows the dependence of the ratio of the apparent speeds of inner
and outer jet components $r_\text{app}=\beta_\text{app, out}/\beta_\text{app, in}$ on the pitch angle for epochs 2008---2010, derived from the expression
\begin{equation}
    r_\text{app}=\frac{\beta_\text{app, out}}{\beta_\text{app, in}}=\frac{\sin \left(\theta\left( \varphi_\text{out}\text{, }p\right) \right)\left(1-\beta \cos \theta \left(\varphi_\text{in}\text{, }p \right) \right)}
    {\sin \left( \theta \left( \varphi_\text{in}\text{, }p \right) \right) \left(1-\beta \cos \theta \left( \varphi_\text{out}\text{, }p\right) \right)},
    \label{eq:rapp}
\end{equation}
where $\varphi_\text{in}=259^\circ$ and $\varphi_\text{out}=101^\circ$.
The observed value of $r_\text{app}\approx2$ \citep{Rani15} is possible only if the jet component speed is $\beta>0.999$ (corresponding to a Lorentz factor $\Gamma \gtrsim 22$). The pitch angle can be equal to
approximately 2$^\circ$, 4.5$^\circ$, 5.5$^\circ$, and 14$^\circ$.
Substituting these values into (\ref{eq:sinthnb})---(\ref{eq:gb}), we found that only for $p = 5.5^\circ$ does the value of $\theta$ vary from 0.5$^\circ$ to 11$^\circ$, corresponding to the interval of estimates of $\theta$, obtained
from apparent superluminal component motions in the jet \citep{Bach05,Rastorgueva11,Pushkarev09,Savolainen10}.
It follows from (\ref{eq:bapp}) that, when $\theta=1^\circ-2^\circ$ and $\beta=0.999-0.9995$, the apparent speed acquires high values $(20c-30c)$, as have
been detected in the jet of S5~0716+71 \citep{Rani15}.
Using VLBI data, \citet{Bach05} estimated the minimum
Lorentz factor to be $\Gamma=11.6$ $(\beta \approx 0.996)$ and the
maximum angle to the line of sight to be $\theta \approx 4.9^\circ$.
However, they note that a large range of observed
apparent component speeds can be realized with $\Gamma>15$ and $\theta<2^\circ$, corresponding to Doppler factors $\delta \approx 20 - 30$.
Precisely such high values of $\delta$ are required
to explain the high brightness temperatures derived
from observations of microvariability of S5~0716+71
at centimeter wavelengths (up to 10$^{17}$~K) \citep{Bach05}.
\citet{Rani13} estimated the Doppler factor using several
independent methods and obtained the lower limit
$\delta\geq10-20$. In their modeling of the spectral energy
distribution in the gamma-ray assuming the action
of inverse Compton scattering on the synchrotron
photons together with inverse Compton scattering on
thermal radiation from hot dust and radiation from
the broad-line region, \citet{Liao14} inferred $\delta$ to
be greater than 20.
\citet{Larionov13} explained the photometric and polarization behavior of a flare in S5~0716+71 in October 2011 as due to the motion
of a radiating region along a helical path with a pitch
angle of $7.5^\circ$ for an angle of the jet to the line of sight
$5.8^\circ$. The long-term trend for increasing brightness
of S5~0716+71 from 1994 through 2003 has been
interpreted as due to a change in the angle of the jet
to the line of sight from $\approx5^\circ$ to 0.7$^\circ$ with a constant $\Gamma=12$ \citep[corresponding to an increase in $\delta$ from 15 to 22,][]{Nesci05}. Thus, the estimates of the jet parameters we have obtained are consistent with other independent
estimates derived using radio, optical, and gamma-ray
observations. 

\begin{figure}
    \centering
    \includegraphics[scale=1.4]{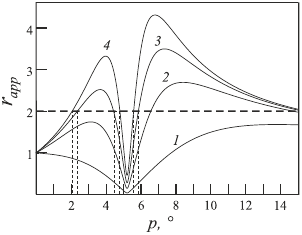}
    \caption{Ratio of the apparent speeds of the outer and
inner jet components $r_\text{app}$ for 2008--2010 as a function
of the pitch angle $p$. Plots are shown for $\beta = 0.995$ (1),
0.999 (2), 0.9995 (3), and 0.9997 (4) (Lorentz factors of
10, 22, 32, and 41, respectively). The horizontal dashed
line shows the observed value of $r_\text{app}$, and the values of $p$
for which this can occur are indicated.}
    \label{fig:fig4}
\end{figure}

\section{conclusion}

The hypothesis that jets have helical structure
has been widely applied. In this case of the blazar
S5~0716+71, such structure is suggested by the variations
of the position angle of the inner jet, which have
a periodic character \citep{Bach05,Lister13}.
In spite of the fact that
the entire interval of VLBI observations of this object
encompasses only slightly more than one calculated
period of variation of PA$_\text{in}$, we have obtained the geometrical parameters of the helical curve of the jet from
the amplitude and period of the variations in PA$_\text{in}$ \citep{Lister13}, namely, the jet opening angle $\xi=1^\circ$ and angle between the axis of the jet helix and the line of sight $\theta_0=5.3^\circ$.
These values are in agreement with analogous
quantities found by \citet{Pushkarev09} based on
VLBI maps, and do not depend on whether or not
the jet components move along ballistic trajectories.
However, with these values of $\theta_0$ and $\xi$ and with non-ballistic motions of the components forming the helical
shape of the jet, the range of values for the angle $\theta$
between the velocity vectors of the components and
the line of sight is appreciably broader than in the
case of ballistic motions --- $\approx10^\circ$, as opposed to $\approx2^\circ$. 
This broad range of $\theta$ is manifest through the different
estimates of $\theta$ obtained in different studies based on
observations at different epochs.
Further, $\theta$ can have very small values ($-1^\circ$), which could result in the high Doppler factors deduced in a number of studies \citep{Montagni06,Bach05}, and also in high apparent speeds for individual components, right up to 37$c$ \cite{Rani15}.
It is also possible to
explain the differences in apparent component speeds
in the inner and outer VLBI jet observed at different
epochs \citep{Bach05,Rastorgueva11} in the case of non-ballistic motions, as well as their different azimuth angles (different
orientations of the components relative to the plane
containing the line of sight and the axis of the helical
curve of the jet).
We have used the observed ratio of $\beta_\text{app}$ in the outer and inner jet to infer the pitch angle $p=5.5^\circ$, at which the components move relative to the ballistic trajectory, and derived a lower limit to the physical speed of the components $\beta>0.999$.
Thus, a helical structure of the jet together with non-ballistic
motions of its components can explain the wide range
of apparent component speeds derived for the inner
and outer jet, as well as the wide range of estimates
of the angle between the velocity vectors of the jet
components and the line of sight.

\acknowledgments{The author thanks A.B. Pushkarev for useful comments.}

\bibliography{helicaljet0716}{}
\bibliographystyle{aasjournal}

\end{document}